# Optimal Choice under Short Sell Limit with Sharpe Ratio as Criterion Among Multiple Assets

Ruokun HUANG*, Yiran SHENG**

**Abstract:**

This article is the term paper of the course *Investments*. We mainly focus on modeling long-term investment decisions of a typical utility-maximizing individual, with features of Chinese stock market in perspective. We adopt an OR based methodology with market information as input parameters to carry out the solution. Two main features of this article are: first, we take the no short-sell constraint in Chinese stock market into consideration and use an approach otherwise identical to Markowitz to work out the optimal portfolio choice; this method has critical and practical implication to Chinese investors. Second, we incorporate the benefits of multiple assets into one single well-defined utility function and use a MIQP procedure to derive the optimal allocation of funds upon each of them along the time-line.

* Ruokun HUANG, 2006012402, huangrk.06@sem.tsinghua.edu.cn, http://learn.tsinghua.edu.cn:8080/2006012402/index

** Yiran SHENG, 2006012400, shengyr.06@sem.tsinghua.edu.cn, http://learn.tsinghua.edu.cn:8080/2006012400/index

# Contents

# I. Background

This article is for a problem described as follows:

- Suppose you are an investor
    - Just graduated
    - 200,000 annual salary
    - 500,000 saving
- How to allocate your assets into:
    - Bank deposit
    - Treasures
    - Stocks
    - Mutual funds
    - Real estate
    - Others

The optimal choice of investment has long been a difficult problem for investors. The structure of people's utility with respect to time and risk influences behaviors of various people. Along with the mainstream method used by scholars, we use well defined utility function to describe investor's attitude toward risk and return.

As we know, there are some existing theories describing the optimal investment choice, i.e. Markowitz method. However, most existing theories are based on relatively complete/perfect market, where short selling of assets is allowed. In our opinion, we must consider this short sell limit in order to avoid discrepancy, which may be large in a market with short sell limit. As a result, we are going to put forward our analysis under short sell limit on stock market.

# II. Introduction

In section III, we show the basic structure of our model and integrate strategy for different investment categories into one universal maximization problem. In section

IV, we further discuss the Monte-Carlo simulation procedure of parameters used in section III. In section V, we give a detailed analysis on stock market with short sell constraint, using an approach otherwise identical to Markowitz. In section VI, we apply some data into our model and compute the results. Section VII concludes.

# III. Our Model-Overall View

To adequately characterize the long-run investment decision of a utility maximizing individual, we need to set up a discrete, multi-period model. We assume that the individual lives for M years, and that his lifetime well-being is determined by the yearend consumption of each period. Furthermore, we assume that he or she has time-additive and state-independent utilities. Thus, the maximization problem can be stated as follows:

$$\max\{E(\sum_{k=1}^{M} U_k(D_k))\}$$

subject to feasible set of $\Theta_k$

Where $D_k$ is his yearend consumption at k^th period, $U_k$ is his corresponding utility in k^th year, and vector $\Theta_k$ is our investment strategy. For simplicity, we assume $U_k$ has uniform structure and a discount factor r along the timeline, that is:

$$U_k = e^{-rk} u(D_k)$$

Next, we will specify the investment strategy $\Theta_k$. We classify our investments into four categories: risky assets, riskless asset (borrow), riskless asset (lend/save), house purchase and insurance product. Denote the money invested on each category in the k^th year is $\theta_{jk}$. Notice insurance is invested only at time 0, we will simplify its purchased amount as $\theta_4$. Here we denote housing expense every year to be $H_k$.

However, since the strike event for insurance happens on an unpredictable date in some future time and may not be exactly the same yearend consumption realization date. Therefore we need to make a little modification in the model, introducing a continuous random variable T called survive time which is the interval starts from time 0 until the strike event happens. We assume T follows some distribution F(t). Once the strike event happens, the person's annual income drops from $I_H$ to $I_L$. The expected utility gain from insurance is, L is the lump sum payment once the strike event happens:

$$V = E(e^{-rT}u(\theta_3 L)) = \int_0^M e^{-ru}u(\theta_3 L)\Pr[T \in du]$$

With this factor included, the full model is:

$$\max_{(\Theta_k, H_k)} \left\{ E\left[ \sum_{k=1}^{M} e^{-rk}u(D_k) + \int_0^M e^{-ru}u\left(\theta_4 L\right)\Pr\left(T \in du\right) + u\left(House\right) \right] \right\}$$

$$D_k = I_L + (I_H - I_L)\delta\left(T > k\right) - \sum_{j=1}^{3}\theta_{jk} + \sum_{j=1}^{3}\theta_{j,k-1}(1+\tilde{R}_j) - s\theta_4\delta\left(T > k\right) - H_k$$

$$\text{where} \quad \delta\left(T > k\right) = \begin{cases} 1, T > k \\ 0, T \le k \end{cases}$$

s.t.

$$D_k \ge \underline{D}$$

$$\theta_{jM} = 0 \text{ ,j=1,2,3}$$

Where $I_H$ and $I_L$ are the given annual income, $\tilde{R}_1$ is the random return on risky portfolio, $\tilde{R}_2 = \overline{r}$ is the risk free (borrow) rate, $\tilde{R}_3 = \underline{r}$ is the risk free (lend/save) rate, s is the yearly spread payment of insurance. $\underline{D}$ is the survival level annual consumption.

Now suppose $E(u(D)) = aE(D) + b\,\mathrm{var}(D)$, the above equation becomes, assuming different category of assets are independent with each other:

$$\max_{(\Theta_k, H_k)} \left\{ \sum_{k=1}^{M} e^{-rk} aE(D_k) + b\,\mathrm{var}(D_k) + V + u\left(House\right) \right\}$$

$$E(D_k) = I_L + (I_H - I_L)(1 - F(k)) - \sum_{j=1}^{3} \theta_{jk} + \sum_{j=1}^{3} \theta_{j,k-1}(1 + E(\tilde{R}_j)) - s\theta_4(1 - F(k))$$

$$\mathrm{var}(D_k) = (I_H - I_L)^2 F(k)(1 - F(k)) + \theta_{1,k-1}{}^2\,\mathrm{var}(\tilde{R}_2) + s^2\theta_4{}^2 F(k)(1 - F(k))$$

The value of V can be derived from Monte-Carlo stimulation, we will discuss this later.

Next, we are going to discuss the term $H_k$. In order to make our planning applicable for computer, we set up a binary vector $\mathbf{B}$, where

$$B_k = \begin{cases} 0, \text{ not buy house at tl} \\ 1, \text{ buy house (pay init} \end{cases}$$

We can easily reach the result that:

$$\mathbf{H} = \begin{pmatrix} Ip & & & & \\ Ap & Ip & & & \\ Ap & Ap & \ddots & & \\ \vdots & \vdots & & \ddots & \\ 0 & & \cdots & Ap & Ip \end{pmatrix} \mathbf{B}$$, $Ip$=Initial payment, $Ap$=Annual payment

We denote the transformation matrix as $\mathbf{P}$

Note that the optimization problem is just a quadratic problem like:

$$\max\left\{\frac{1}{2}\mathbf{x}^{\mathbf{T}}\mathbf{\Sigma x} + \mathbf{c}^{\mathbf{T}}\mathbf{x}\right\}$$
$$s.t.\mathbf{Ax} \geq \mathbf{b}$$

with $\mathbf{x} = \left(\theta_{1,1}, \theta_{1,2} \ldots \theta_{1,M}, \theta_{2,1} \ldots \theta_{2,M}, \theta_{3,1} \ldots \theta_{3,M}, H_1 \ldots H_M, \theta_4\right)^T$

This problem is equivalent to:

$$\max\left\{\frac{1}{2}\mathbf{y}^{\mathbf{T}}\mathbf{\Sigma y} + \mathbf{d}^{\mathbf{T}}\mathbf{y}\right\}$$
$$s.t.\mathbf{Dx} \geq \mathbf{e}$$

with $\mathbf{y} = \left(\theta_{1,1}, \theta_{1,2} \ldots \theta_{1,M}, \theta_{2,1} \ldots \theta_{2,M}, \theta_{3,1} \ldots \theta_{3,M}, B_1 \ldots B_M, \theta_4\right)^{\mathbf{T}}$

and $\mathbf{d}^{\mathrm{T}} = \mathbf{c}^{\mathrm{T}} \left( \begin{array}{ccc|c|c} 1 & & 0 & 0 & 0 \\ & \ddots & & \vdots & \vdots \\ 0 & & 1 & 0 & 0 \\ \hline 0 & \cdots & 0 & \mathbf{P} & 0 \\ \hline 0 & \cdots & 0 & 0 & 1 \end{array} \right)$

This is always applicable since $|\mathbf{P}| \neq 0$.

# IV. Determination of Insurance value via Monte-Carlo Stimulation

In order to specify the mechanism of insurance product in a sense of its contribution to life-time utilities, we introduce a latent variable model. Suppose Y is a latent variable follows standard normal distribution. When Y falls below a critical value, the strike event in insurance contract happens, *id est*:

$$\Pr[T \le t] = F(t) = \Pr[Y \le y] = \Phi(y)$$

Therefore there is a one-to-one mapping between Y and T:

$$T = F^{-1}(\Phi(Y))$$

Since T is a rv in nature, we need to carry out a Monte-Carlo Stimulation to determine the expected utility gain from insurance products. Throughout this session we assume $F(t)=1-e^{-ht}$; where h is a constant hazard rate, its mathematical meaning is the strike event conditional probability density at any time.

$$T = -\frac{\log(1-\Phi(Y))}{h}$$

The following table gives input parameters for the stimulation:

| Parameters | Input value |
|---|---|
| Hazard rate: h | 0.06 |
| Lump sum payments: L | 30000 |
| Spread payment: s | 500 |

For h =0.06, the probability that the strike event will not happen within 30 years is about 17%.

After 10000 stimulation runs over Y, we have V = 0.6629u($\theta_4$L)

# V. Our Model-Stock Market Optimal Choice under Short Sell Limit

Firstly, in classical model, where short selling of assets is allowed, the optimal problem has simple solution as follows:

$$\mathbf{w} = \frac{\mathbf{\Sigma}^{-1}\mathbf{e}\left[E\left(\tilde{r}_M\right) - r_f\right]}{H}$$

Where $E\left(\tilde{r}_M\right) = \frac{A}{C} - \frac{D}{C^2\left(r_f - \frac{A}{C}\right)}$,

$$A = \mathbf{1}^{\mathbf{T}}\mathbf{\Sigma}^{\mathbf{-1}}\mathbf{e}$$

$$B = \mathbf{e}^{\mathbf{T}}\mathbf{\Sigma}^{\mathbf{-1}}\mathbf{e}$$

$$C = \mathbf{1}^{\mathbf{T}}\mathbf{\Sigma}^{\mathbf{-1}}\mathbf{1}$$

$$D = BC - A^2$$

$$H = B - 2Ar_f + C{r_f}^2$$

Actually under Markowitz's world, it is the solution of the maximizing Sharpe Ratio. Here in our analysis, we also use Sharpe Ratio as a criterion of portfolio evaluation. However, we assume that short sell is limited, as in Chinese stock market.

The mathematical description of our analysis is as follows:

$$\max_{\mathrm{w}}\left\{\frac{E\left[\tilde{r}_{\mathrm{w}}\right] - r_f}{\sigma_{\mathrm{w}}}\right\}$$

subject to $\begin{cases} \mathbf{1}^{\mathbf{T}}\mathbf{w} = 1, \\ w_i \geq 0, \forall i \end{cases}$

There is no simple solution to this non-linear planning problem as that of Markowitz. Hereby we use software R (R Development Core Team, 2008), package

Rdonlp2 to solve this problem.

Secondly, there are too many stocks in the market, specifically more than 1600 in A share market. From a general view these stocks are to some extent homogeneous, since they are often highly related. In our analysis, for the convenience of coumputing, we choose several stocks from the same industry, *id est* 深发展 A (000001) from Financial Institution industry, *et cetera*. Hereby we choose 143 stocks from the 1600+ stocks in A share market, listed in Appendix A (page 12).

# VI. Computing Process and Numerical Result

The data we choose are as follows:

Stock: 143 A stocks as in Appendix A (page 12).

Utility discount rate: 3%

Risk free rate (Borrow): 6.5%

Risk free rate (Lend/save):2.5%

House initial payment: $1,800,000

House annual payment: $150,000

(other input data are in Appendix B-R code)

The numerical result is:

Choose these stocks as a *mutual fund*.

| 变量 No. | 股票号码 | 股票名称 | 比例 |
|---|---|---|---|
| 2 | 000002 | 万科 A | 0.142771581 |
| 13 | 000400 | 许继电气 | 0.0105711138 |
| 22 | 000515 | 攀渝钛业 | 0.0230618999 |
| 25 | 000538 | 云南白药 | 0.2125265671 |
| 35 | 000661 | 长春高新 | 0.0240573083 |
| 40 | 000792 | 盐湖钾肥 | 0.2222784446 |
| 46 | 000816 | 江淮动力 | 0.0147827086 |
| 53 | 000895 | 双汇发展 | 0.0642883098 |
| 75 | 600038 | 哈飞股份 | 0.010492248 |
| 83 | 600096 | 云天化 | 0.070783097 |
| 113 | 600519 | 贵州茅台 | 0.1458883638 |
| 139 | 600875 | 东方电气 | 0.058498358 |

Invest in stock market by buying following pieces of *mutual fund*:

| Year | Amount | Year | Amount |
|---|---|---|---|
| 1 | 0.02688647 | 16 | 0.00929332 |
| 2 | 0.02524552 | 17 | 0.00901866 |
| 3 | 0.02462977 | 18 | 0.00875212 |
| 4 | 0.02402905 | 19 | 0.00849345 |
| 5 | 0.02344297 | 20 | 0.00824243 |
| 6 | -0.00029566 | 21 | 0.00799883 |
| 7 | 0.01273575 | 22 | 0.00776243 |
| 8 | 0.01236226 | 23 | 0.00753302 |
| 9 | 0.01146496 | 24 | 0.00731038 |
| 10 | 0.01112612 | 25 | 0.00709433 |
| 11 | 0.0107973 | 26 | 0.00688466 |
| 12 | 0.01047819 | 27 | 0.00668119 |
| 13 | 0.01016851 | 28 | 0.00648373 |
| 14 | 0.009867986 | 29 | 0.00629211 |
| 15 | 0.009576343 | 30 | 0 |

The borrowing behavior is:

Borrow 25.41707 pieces= $25,417.07 at year 6, zero otherwise.

The saving behavior is: (zero otherwise)

| Year | Amount |
|---|---|
| 1 | $ 689,222.7 |
| 2 | $ 895,705.3 |
| 3 | $1,107,349 |
| 4 | $1,324,284 |
| 5 | $1,546,643 |

Buy house at year 6.

Besides, the *mutual fund* has the frontier like:

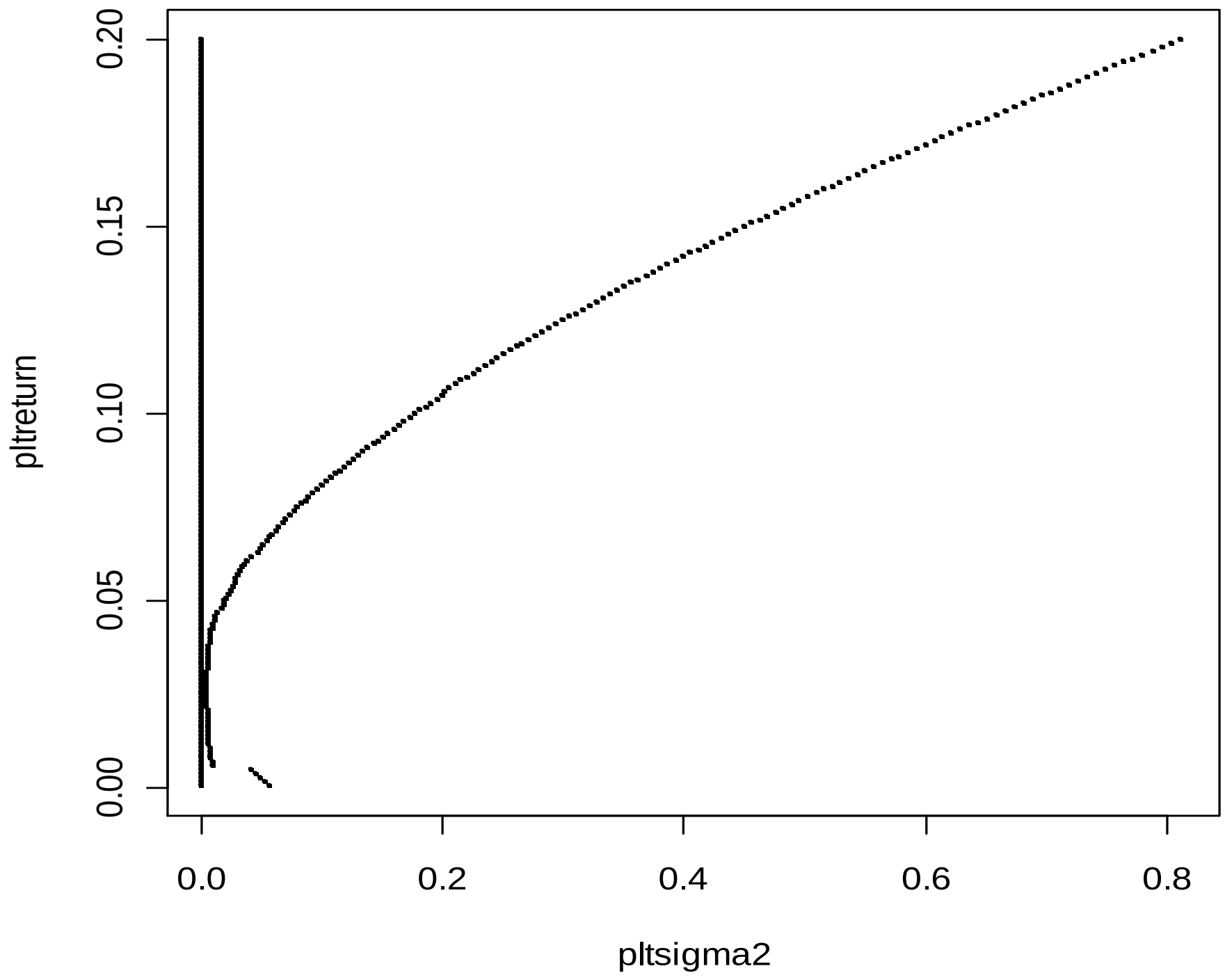

The 'straight line' in the graph above is the original frontier without short sell limit, which appears to be much larger than our frontier. Note that that frontier is a parabolic curve at a proper scale.

# VII. Conclusion

**a) Short sell limit plays an important role in Chinese stock market**

From the graph we showed above, we can see that with limit of short sell, the feasible set shrinks much compared to situation without short sell constraint. So, we must consider short sell limit in Chinese stock market.

**b) Stock market as an asset of investment should be careful considered**

In our optimal solution, *mutual fund* is invested at relatively a very low level. This implies one should better choose other investment assets other than stock.

**c) House purchase is the core of life due to large amount of utility it brings**

All the financing method in our analysis, including borrowing and saving, contribute to the 'final target' of buying a house. The optimal choice indicates one should save money for 5 years, and as long as he has enough or near to enough money, he purchases a house with small amount of loan. This solution is very close to modern Chinese young people.

**d) Further work to be done**

The overall utility assumption also has its limitations. The state independent assumption tends to be invalid since the insurance is introduced in this model. The utility of house can be decided by more factors. Besides, the period of house payment and/or payment structure should be flexible.

# Appendix A - Stock to be Selected

| No | 股票号码 | 股票名称 | No | 股票号码 | 股票名称 | No | 股票号码 | 股票名称 |
|---|---|---|---|---|---|---|---|---|
| 1 | 000001 | 深发展 A | 51 | 000860 | 顺鑫农业 | 101 | 600299 | 蓝星新材 |
| 2 | 000002 | 万科 A | 52 | 000878 | 云南铜业 | 102 | 600300 | 维维股份 |
| 3 | 000009 | 中国宝安 | 53 | 000895 | 双汇发展 | 103 | 600305 | 恒顺醋业 |
| 4 | 000012 | 南玻 A | 54 | 000911 | 南宁糖业 | 104 | 600313 | ST 中农 |
| 5 | 000031 | 中粮地产 | 55 | 000931 | 中关村 | 105 | 600315 | 上海家化 |
| 6 | 000034 | ST 深泰 | 56 | 000933 | 神火股份 | 106 | 600320 | 振华港机 |
| 7 | 000040 | 深鸿基 | 57 | 000936 | 华西村 | 107 | 600333 | 长春燃气 |
| 8 | 000043 | 中航地产 | 58 | 000938 | 紫光股份 | 108 | 600350 | 山东高速 |
| 9 | 000049 | 德赛电池 | 59 | 000951 | 中国重汽 | 109 | 600356 | 恒丰纸业 |
| 10 | 000060 | 中金岭南 | 60 | 000962 | 东方钽业 | 110 | 600382 | 广东明珠 |
| 11 | 000061 | 农产品 | 61 | 000968 | 煤气化 | 111 | 600416 | 湘电股份 |
| 12 | 000089 | 深圳机场 | 62 | 000972 | 新中基 | 112 | 600418 | 江淮汽车 |
| 13 | 000400 | 许继电气 | 63 | 000988 | 华工科技 | 113 | 600519 | 贵州茅台 |
| 14 | 000401 | 冀东水泥 | 64 | 000990 | 诚志股份 | 114 | 600528 | 中铁二局 |
| 15 | 000402 | 金融街 | 65 | 000998 | 隆平高科 | 115 | 600530 | 交大昂立 |
| 16 | 000410 | 沈阳机床 | 66 | 000999 | 三九医药 | 116 | 600559 | 老白干酒 |
| 17 | 000420 | 吉林化纤 | 67 | 600001 | 邯郸钢铁 | 117 | 600585 | 海螺水泥 |
| 18 | 000423 | 东阿阿胶 | 68 | 600007 | 中国国贸 | 118 | 600587 | 新华医疗 |
| 19 | 000425 | 徐工科技 | 69 | 600008 | 首创股份 | 119 | 600597 | 光明乳业 |
| 20 | 000426 | 富龙热电 | 70 | 600011 | 华能国际 | 120 | 600598 | 北大荒 |
| 21 | 000509 | SST 华塑 | 71 | 600026 | 中海发展 | 121 | 600611 | 大众交通 |
| 22 | 000515 | 攀渝钛业 | 72 | 600028 | 中国石化 | 122 | 600631 | 百联股份 |
| 23 | 000518 | 四环生物 | 73 | 600036 | 招商银行 | 123 | 600633 | *ST 白猫 |
| 24 | 000527 | 美的电器 | 74 | 600037 | 歌华有线 | 124 | 600655 | 豫园商城 |
| 25 | 000538 | 云南白药 | 75 | 600038 | 哈飞股份 | 125 | 600661 | 交大南洋 |
| 26 | 000552 | 靖远煤电 | 76 | 600062 | 双鹤药业 | 126 | 600663 | 陆家嘴 |
| 27 | 000554 | 泰山石油 | 77 | 600064 | 南京高科 | 127 | 600701 | 工大高新 |
| 28 | 000585 | 东北电气 | 78 | 600066 | 宇通客车 | 128 | 600710 | 常林股份 |
| 29 | 000591 | 桐君阁 | 79 | 600073 | 上海梅林 | 129 | 600723 | 西单商场 |
| 30 | 000597 | 东北制药 | 80 | 600085 | 同仁堂 | 130 | 600737 | 中粮屯河 |
| 31 | 000598 | 蓝星清洗 | 81 | 600087 | 长航油运 | 131 | 600754 | 锦江股份 |
| 32 | 000619 | 海螺型材 | 82 | 600090 | 啤酒花 | 132 | 600756 | 浪潮软件 |
| 33 | 000629 | 攀钢钢钒 | 83 | 600096 | 云天化 | 133 | 600775 | 南京熊猫 |
| 34 | 000630 | 铜陵有色 | 84 | 600100 | 同方股份 | 134 | 600798 | 宁波海运 |
| 35 | 000661 | 长春高新 | 85 | 600106 | 重庆路桥 | 135 | 600806 | 昆明机床 |
| 36 | 000682 | 东方电子 | 86 | 600109 | 国金证券 | 136 | 600835 | 上海机电 |
| 37 | 000729 | 燕京啤酒 | 87 | 600111 | 包钢稀土 | 137 | 600849 | 上海医药 |
| 38 | 000758 | 中色股份 | 88 | 600115 | 东方航空 | 138 | 600867 | 通化东宝 |
| 39 | 000768 | 西飞国际 | 89 | 600135 | 乐凯胶片 | 139 | 600875 | 东方电气 |
| 40 | 000792 | 盐湖钾肥 | 90 | 600138 | 中青旅 | 140 | 600879 | 火箭股份 |
| 41 | 000798 | 中水渔业 | 91 | 600150 | 中国船舶 | 141 | 600887 | 伊利股份 |
| 42 | 000799 | 酒鬼酒 | 92 | 600162 | 香江控股 | 142 | 600889 | 南京化纤 |
| 43 | 000802 | 北京旅游 | 93 | 600169 | 太原重工 | 143 | 600895 | 张江高科 |
| 44 | 000807 | 云铝股份 | 94 | 600186 | 莲花味精 | | | |
| 45 | 000811 | 烟台冰轮 | 95 | 600188 | 兖州煤业 | | | |
| 46 | 000816 | 江淮动力 | 96 | 600192 | 长城电工 | | | |
| 47 | 000830 | 鲁西化工 | 97 | 600195 | 中牧股份 | | | |
| 48 | 000833 | 贵糖股份 | 98 | 600197 | 伊力特 | | | |
| 49 | 000856 | 唐山陶瓷 | 99 | 600229 | 青岛碱业 | | | |
| 50 | 000858 | 五粮液 | 100 | 600283 | 钱江水利 | | | |

# Appendix B - R code

```
library("Rcplex")
library("Matrix")
library("xlsReadWrite")
library("Rdonlp2")
#Read libraries
#Parameters, '1'=$1000
MonteCarloNumber=10000       #Times of experiment in Monte-Carlo
InitialSaving=500            #Initial saving
years=30                     #Total Years
HouseInitial=1800            #House Initial payment
HouseAnnual=150              #House annual payment
HouseY=10                    #House payment years
r=0.03                       #Utility Discount rate
```

```
rbottom=0.025               #Risk free (lend,save) rate
rtop=0.065                  #Risk free (Borrow) rate
L=30                        #Insurance lamp-sum payment
HouseUtil=3500              #utility of House, in money sense
h=0.06                      #hazard rate
s=.5                        #spread of Insurance
Dbottom=10                  #lowest living expense
Ih=200                      #Annual Salary
Il=10                       #Annual salary after losing job
B=3                         #risk averse
rhouse=0.0                  #House price growth rate
yy=rnorm(MonteCarloNumber,0,1)
tt=-log(1-pnorm(yy))/h
vv=exp(-r*tt)
V=mean(vv)                  #Calculate V using Monte Carlo
kstart=ceiling(mean(tt))
Fk<-function(x){
   tst=1-exp(-x*h)
   tst
}#Define Fk
#start to optimize stock
fn <- function(x){
 -(expectret%*%x - rbottom/12)/ sqrt(+t(x)%*%sigma%*%x)  #Min fn.}
inmat=read.xls("s2ok.xls",sheet=1,type="double")
n=ncol(inmat)
sigma=cov(inmat);expectret=apply(inmat,MARGIN=2,FUN=mean)
p <- rep((1/n),n)           #initial value
par.l <- rep(0,n); par.u <- rep(1,n)   #Parameter range
lin.u <- 1; lin.l <- 1
A <- t(rep(1,n))
ret <- donlp2(p, fn, par.lower=par.l, par.upper=par.u,
            A=A, lin.u=lin.u, lin.l=lin.l, name="stockopl")
rstock<-expectret%*%(ret$par)
sigma2stock=t(ret$par)%*%sigma%*%(ret$par)*12
sigma2stock=as.numeric(sigma2stock)
rstock=as.numeric(rstock)
#vector=(stock1,stock2...stock30,borrow1,borrow2..borrow30,save1...sa
ve30,B1...B30,Insurance)
tempm=diag(HouseInitial,years)
for (i in 1:years){
   for (j in (i+1):(i+HouseY)){
       if (j<=years ) tempm[j,i]=HouseAnnual  }
   tempm[i,i]=HouseInitial*exp(i*rhouse)}
transform_mat_q=as.matrix(bdiag(diag(1,3*years),tempm,1))
```

```
term1_v1=matrix(0,1,4*years+1)
for (i in 1:years){
   if (i<years) term1_v1[i]=-exp(-i*r)+exp(-(i+1)*r)*(1+rstock) else
term1_v1[i]=-exp(-i*r)
}#coefficient of stock
for (i in 1:years){
   if  (i<years)  term1_v1[i+years]=+exp(-i*r)-exp(-(i+1)*r)*(1+rtop)
else term1_v1[i+years]=-exp(-i*r)
}#coefficient of borrow
for (i in 1:years){
   if (i<years)
term1_v1[i+2*years]=-exp(-i*r)+exp(-(i+1)*r)*(1+rbottom) else
term1_v1[i+2*years]=-exp(-i*r)
}#coefficient of save
for (i in 1:years){
   term1_v1[i+3*years]=-exp(-i*r)
}#coefficient of house
mysum=0
for (i in 1:years){mysum=mysum+1-Fk(i)}
term1_v1[1+4*years]=V*L-s*mysum
term1_v2=matrix(0,1,4*years+1)
for (i in 1:years){
   term1_v2[i+3*years]=exp(-i*r)*HouseUtil}
term1=term1_v1%*%transform_mat_q+term1_v2
myc=t(term1)
#first order ok.
mysum=0
for (i in 1:years){mysum=mysum+(1-Fk(i))*Fk(i)}
Q=as.matrix(bdiag(diag(sigma2stock,years),diag(0,3*years),s*s*mysum))
#second order ok.
a1=diag(-1,years)
   for (i in 2:years){  a1[i,i-1]=1+rstock}
a3=diag(-1,years)
   for (i in 2:years){  a3[i,i-1]=1+rbottom}
a2=diag(1,years)
   for (i in 2:years){  a2[i,i-1]=-1-rtop}
a4=diag(-1,years); a4=a4%*%tempm
a5=rep(-s,years)
   for (i in kstart:years){a5[i]=0}
a=cbind(a1,a2,a3,a4,a5)
al=t(rep(0,4*years+1))
   for (i in (3*years+1):(4*years)){al[i]=-1}
a=rbind(a,al)
#Amat ok
```

```
b=rep(Dbottom-Ih,years+1)
   for (i in kstart:years){b[i]=Dbottom-Il}
b[years+1]=-1
b[1]=b[1]-InitialSaving
#bvec ok.
Q=Q*(-2)*B
myvtype=rep("C",4*years+1)
for (i in (3*years+1):(4*years)){myvtype[i]="B"}
ub=rep(Inf,4*years+1)
for (i in (4*years-HouseY+1):(4*years)){ub[i]=0}
myresult<-
Rcplex(cvec=myc,Amat=a,bvec=b,Qmat=Q,lb=0,ub=ub,objsense="max",
sense="G" , vtype=myvtype)
print(myresult)
```

# Appendix C – R result

The optimal choice of stocks under short sell limit, whose parameter is not zero:

| 变量No. | 股票号码 | 股票名称 | 比例 |
|---|---|---|---|
| 2 | 000002 | 万科A | 0.142771581 |
| 13 | 000400 | 许继电气 | 0.0105711138 |
| 22 | 000515 | 攀渝钛业 | 0.0230618999 |
| 25 | 000538 | 云南白药 | 0.2125265671 |
| 35 | 000661 | 长春高新 | 0.0240573083 |
| 40 | 000792 | 盐湖钾肥 | 0.2222784446 |
| 46 | 000816 | 江淮动力 | 0.0147827086 |
| 53 | 000895 | 双汇发展 | 0.0642883098 |
| 75 | 600038 | 哈飞股份 | 0.010492248 |
| 83 | 600096 | 云天化 | 0.070783097 |
| 113 | 600519 | 贵州茅台 | 0.1458883638 |
| 139 | 600875 | 东方电气 | 0.058498358 |

Final MIQP result is: (total years=30)

```
  [1]  2.688647e-02  2.524552e-02  2.462977e-02  2.402905e-02  2.344297e-02
  [6] -2.956562e-04  1.273575e-02  1.236226e-02  1.146496e-02  1.112612e-02
 [11]  1.079730e-02  1.047819e-02  1.016851e-02  9.867986e-03  9.576343e-03
 [16]  9.293319e-03  9.018660e-03  8.752118e-03  8.493454e-03  8.242434e-03
 [21]  7.998834e-03  7.762432e-03  7.533018e-03  7.310384e-03  7.094329e-03
 [26]  6.884660e-03  6.681188e-03  6.483729e-03  6.292105e-03  0.000000e+00
 [31]  0.000000e+00 -3.409720e-04  0.000000e+00  0.000000e+00  0.000000e+00
 [36]  2.541707e+01  0.000000e+00  0.000000e+00  0.000000e+00  0.000000e+00
 [41]  0.000000e+00  0.000000e+00  0.000000e+00  0.000000e+00  0.000000e+00
 [46]  0.000000e+00  0.000000e+00  0.000000e+00  0.000000e+00  0.000000e+00
 [51]  0.000000e+00  0.000000e+00  0.000000e+00  0.000000e+00  0.000000e+00
```

```
 [56]  0.000000e+00  0.000000e+00  0.000000e+00  0.000000e+00  0.000000e+00
 [61]  6.892227e+02  8.957053e+02  1.107349e+03  1.324284e+03  1.546643e+03
 [66]  0.000000e+00  0.000000e+00  0.000000e+00  0.000000e+00  0.000000e+00
 [71]  0.000000e+00  0.000000e+00  0.000000e+00  0.000000e+00  0.000000e+00
 [76]  0.000000e+00  0.000000e+00  0.000000e+00  0.000000e+00  0.000000e+00
 [81]  0.000000e+00  0.000000e+00  0.000000e+00  0.000000e+00  0.000000e+00
 [86]  0.000000e+00  0.000000e+00  0.000000e+00  0.000000e+00  0.000000e+00
 [91]  0.000000e+00  0.000000e+00  0.000000e+00  0.000000e+00  0.000000e+00
 [96]  1.000000e+00  0.000000e+00  0.000000e+00  0.000000e+00  0.000000e+00
[101]  0.000000e+00  0.000000e+00  0.000000e+00  0.000000e+00  0.000000e+00
[106]  0.000000e+00  0.000000e+00  0.000000e+00  0.000000e+00  0.000000e+00
[111]  0.000000e+00  0.000000e+00  0.000000e+00  0.000000e+00  0.000000e+00
[116]  0.000000e+00  0.000000e+00  0.000000e+00  0.000000e+00  0.000000e+00
[121]  1.500728e+00
```